\documentclass[a4paper,11pt]{article}
\usepackage{pos}
\usepackage{physics}

\usepackage{color}

\title{Study on Lambda(1405) in the flavor SU(3) limit in the HAL QCD method}

\author*[a,b]{Kotaro Murakami}
\author[c]{Sinya Aoki}

\affiliation[a]{Department of Physics, Tokyo Institute of Technology, \\ 
2-12-1 Ookayama, Megro, Tokyo 152-8551, Japan}

\affiliation[b]{Interdisciplinary Theoretical and Mathematical Sciences Program (iTHEMS), RIKEN, Wako 351-0198, Japan}

\affiliation[c]{Center for Gravitational Physics, Yukawa Institute for Theoretical Physics, Kyoto University, \\
Kitashirakawa Oiwakecho, Sakyo-ku, Kyoto 606-8502, Japan}

\emailAdd{kotaro.murakami@yukawa.kyoto-u.ac.jp}
\emailAdd{saoki@yukawa.kyoto-u.ac.jp}

\abstract{
We study interactions between the S-wave octet pseudo-scalar (PS) meson and octet baryon in the flavor SU(3) limit using the HAL QCD method at the PS meson mass $m_M\approx 670~\textrm{MeV}$.
We focus on the singlet and two octet channels, where the poles corresponding to $\Lambda(1405)$ have been predicted in the chiral unitary model. 
For calculations with $\Lambda$-baryon source operators with zero momentum, we employ the conventional stochastic calculation combined with the covariant-approximation averaging to calculate the all-to-all propagators.
Due to a zero of the R-correlator (a kind of wave function), 
the leading order (LO) potential obtained by the single channel analysis has a singular point in all channels, which makes it difficult to obtain reliable binding energies. 
To overcome this problem, we take a  linear combination of two octet R-correlators with a relative weight such that it does not cross zero, as two octet channels are suggested to couple to the same low-energy states with different weights.
The potential calculated from such the linear combination shows strong attraction without singularities, though its shape depends on the relative weight. 
Our estimation for the binding energy in the octet channel is  $E^{8_{s(a)}}_{\textrm{bind}}=163(7)\smqty(+16 \\ -64)~\textrm{MeV}$, which is consistent with 
156(8) MeV estimated from the two-point correlation function  within  errors.
}

\FullConference{The 40th International Symposium on Lattice Field Theory (Lattice 2023)\\
July 31st - August 4th, 2023\\
Fermi National Accelerator Laboratory\\}

\begin{document}
\maketitle

\section{Introduction}
\label{sec:intro}

Understanding exotic hadrons through scattering processes derived in lattice QCD is one of the essential issues in hadron physics. 
One of the methods to study hadron scatterings is the HAL QCD method~\cite{Ishii:2006ec,Aoki:2009ji,Ishii:2012ssm}, which extracts interaction potentials between hadrons. 
The new all-to-all calculation technique recently proposed~\cite{Akahoshi:2021sxc} makes HAL QCD studies on hadron resonances and exotic hadrons with quark-pair annihilation diagrams possible. 

In this paper, we investigate $\Lambda(1405)$, a resonance with $S=-1$, $I=0$ and $J^{P}=1/2^{-}$, observed below $N\bar{K}$ and above $\Sigma\pi$ thresholds. 
This hadron cannot be explained by a simple 3-quark state in the quark model. 
Although numerous theoretical and experimental studies have been made, its internal structure and properties are still controversial. 
One of the plausible explanations for $\Lambda(1405)$ proposed in the chiral unitary model~\cite{Oller:2000fj, Jido:2003cb}. 
is that its behavior is controlled by two poles in the scattering amplitude. 
Our goal in this study is to examine the existence and the mechanism of such a two-pole structure by the HAL QCD method.

As a first step toward the goal, we investigate $\Lambda(1405)$ in the flavor SU(3) limit, where the analysis is much simpler than that in the $2+1$-flavor case. 
The previous study by the chiral unitary model~\cite{Jido:2003cb} suggests that the two poles constructing $\Lambda(1405)$ correspond to one in the singlet channel and the other in the octet channels in the SU(3) limit.
In this study, we investigate S-wave interaction potentials between  PS octet meson and octet baryon  in the HAL QCD method to see the existence of poles in these channels.

\section{Time-dependent HAL QCD method in meson-baryon systems}
\label{sec:HAL}
Let us start with the R-correlator defined by 
\begin{eqnarray}
R_{\alpha}(\vb{r},t)  
=\frac{\langle M(\vb{r+x},t+t_{0})B_{\alpha}(\vb{x},t+t_{0}) \bar{J}(t_0) \rangle}{C_{M}(t)C_{B}(t)},
\end{eqnarray}
where $M(\vb{x},t)$ and $B_{\alpha}(\vb{x},t)$ are operators located at $(\vb{x},t)$ for meson and baryon, respectively, $\bar{J}(t_0)$ is the source operator of the meson-baryon states at the time $t_0$, and
$C_{M}(t)$ and  $C_{B}(t)$ are  two-point correlation functions of meson and baryon, respectively. 
The numerator on the right-hand side is called the $n$-point correlation function with $n\geq3$.
Using the fact that the R-correlator at large $t$ can be expressed by a linear combination of equal-time Nambu-Bethe-Salpeter (NBS) wave functions, we obtain the equation for an interaction potential, $U(\vb{r},\vb{r}')$ with the R-correlator at large $t$ as
\begin{eqnarray}\label{eq:scheq_rela_approx}
\int d^3r' \ U_{\alpha\beta}(\vb{r},\vb{r}')R_{\beta}(\vb{r'},t) 
\simeq
\left(\frac{\nabla^2}{2\mu}-\pdv{t} + \frac{1+3\delta^2}{8\mu}\pdv[2]{t} \right)R_{\alpha}(\vb{r},t) + \order{\Delta W^3},
\end{eqnarray}
where $\mu$ is the reduced mass, $\delta=(m_M-m_B)/(m_M+m_B)$, and $\Delta W$ is the typical energy of the meson-baryon system from the threshold, $\Delta W=W-m_M-m_B$. 
Then,LO potential $V^{LO}(r)$ with $U(\vb{r},\vb{r}')\simeq V^{LO}(r) \delta^{(3)}(\vb{r}-\vb{r}')$ can be computed from a single R-correlator as 
\begin{eqnarray}\label{eq:LOpotential_gen}
V^{\textrm{LO}}(r) \simeq
\frac{1}{R_{\alpha}(\vb{r},t)}\left(\frac{\nabla^2}{2\mu}-\pdv{t} + \frac{1+3\delta^2}{8\mu}\pdv[2]{t} \right)R_{\alpha}(\vb{r},t).
\end{eqnarray}
We then  solve the Schr\"{o}dinger equation with the obtained potential to extract scattering phase shifts and  binding energies.

\section{Setups}
\label{sec:setups}
In the present calculation, we use gauge configurations in flavor-SU(3) limit with the improved Iwasaki gauge action and the $\order{a}$-improved Wilson quark action at $\beta = 1.83$ on $32^4$ lattice volume~\cite{Inoue:2015dmi},
whose lattice spacing $a=0.121(2)~\textrm{fm}$ and hopping parameters are $\kappa_{u}=\kappa_{d}=\kappa_{s}=0.13800$.
We employ 360 configurations and 32 smeared quark sources in Ref.~\cite{Iritani:2016jie} with $(A,B)=(1.2, 0.3)$
at different time slices on each configuration. 
The same smearing is imposed at the sink with $(A,B)=(1.0, 1/0.7)$ to reduce the non-smoothness of potentials at short distances~\cite{Akahoshi:2021sxc}. 

Hadron masses extracted from two-point correlation functions are listed in Table.~\ref{tab:mass_list}, where we denote the PS meson mass as $m_M$ and the baryon mass with the parity $P$ and the representation $R$ as $m^{P}_{R,B}$.
We find that there is at least one bound state for each channel in this setup since the inequalities $m^{-}_{8,B}-m^{+}_{8,B}-m_{M}<0$ and $m^{-}_{1,B}-m^{+}_{8,B}-m_{M}<0$ are satisfied. 

\begin{table}
    \centering     
    \begin{tabular}{c|c|c|c}
     $m_{M}$      & 
    $m^{+}_{8,B}$ &
    $m^{-}_{8,B}$ &
    $m^{-}_{1,B}$
     \\ \hline
     671.2(1.5)&
    1488.8(3.9)&
    2013.2(11.2)&
    1923.4(9.1)
     \\ \hline
     [8,15]&
    [8,13] &
    [6,11] &
    [5,11]
    \end{tabular}        
    \caption{\label{tab:mass_list}
    Pseudo-scalar meson and baryon masses in MeV estimated from the 2-point correlation functions.
    The third row shows temporal fitting ranges in lattice unit.
    }
\end{table}

In this study, we focus on the singlet and the two octet channels, $8_{s}$ and $8_{a}$, which contain two poles corresponding to $\Lambda(1405)$ in the chiral unitary model, where
$8_{s}$ ($8_{a}$) is defined to be symmetric (anti-symmetric) by exchanging flavors in meson and baryon.
We calculate the three-point correlation function using meson-baryon sink operators in the three representations, 
and the singlet $\Lambda$-baryon source operator with negative parity projected onto zero momentum for the singlet channel
as well as the octet one for $8_{s}$ and $8_{a}$ channels. 
In our calculation, all-to-all propagators are necessary since the spatial coordinate at the source is summed over. 
To calculate them, we use the conventional stochastic technique combined with the dilutions~\cite{Foley:2005ac} for color/spinor/time components and the s4 dilution for spaces~\cite{Akahoshi:2019klc}. 
We also employ the covariant-approximation averaging~\cite{Shintani:2014vja} combined with the truncated solver method~\cite{Bali:2009hu} using the translational invariance of the baryon operator at the sink. 
Finally, to obtain the S-wave component of the LO potential, we project each spin of three-point correlation functions onto the A$^{+}_{1}$ representation and take an average over spins. 

In general, $8_{s}$ and $8_{a}$ channels can couple to each other, so that the coupled-channel analysis is necessary. 
In this study, we ignore the coupling and perform the single-channel analysis in each channel. 
Although such an effect on the observables should be examined in the future, this approximation is justfied by the chiral perturbation theory with the Weinberg-Tomozawa interaction~\cite{Jido:2003cb}, where $8_s$ and $8_a$ do not mix.


\section{Numerical results}
\label{sec:numerical_results}
\subsection{Leading-order potential in each representation}
\label{subsec:results_orig}

\begin{figure}[h]
    \begin{center}
        \includegraphics[width=0.45\textwidth]{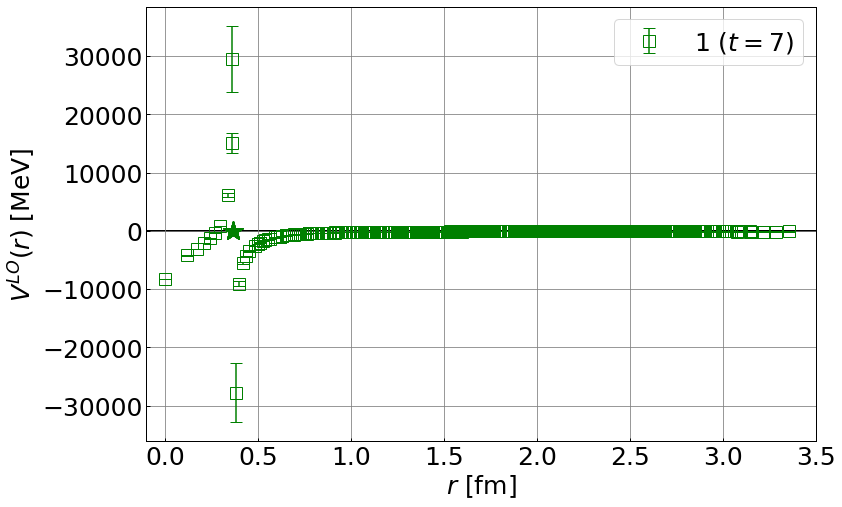}
        \includegraphics[width=0.45\textwidth]{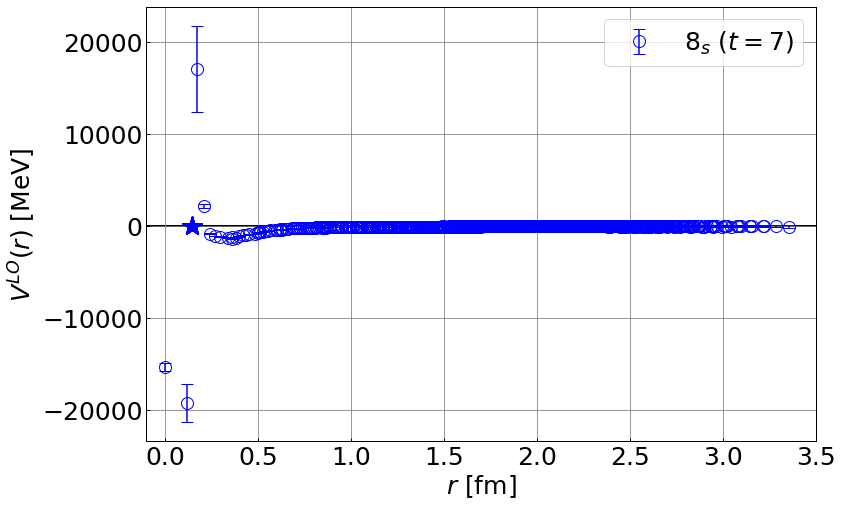}
        \includegraphics[width=0.45\textwidth]{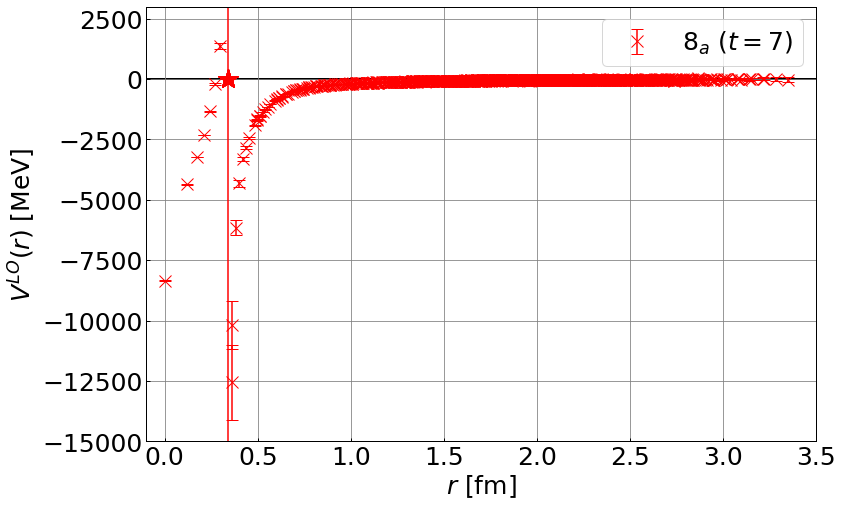}
	    \caption{LO potentials in $1$ (Upper left panel), $8_s$ (Upper right panel), and $8_a$ chennels (Lower panel) at $t=7$. 
        Star symbols on horizontal axes correspond to zeros of R-correlators. 
        }
	\label{fig:results_pot_orig}    
    \end{center}
\end{figure}

We first show the LO potentials in Fig.~\ref{fig:results_pot_orig}. 
We find a singular behavior in the potential for all channels; it has a quite large value and seems to be singular at some $r$. 
This occurs  because the R-correlator in the denominator of Eq.~\eqref{eq:LOpotential_gen} vanishes at the corresponding $r$. 
Indeed, as seen in the figure, the zero of the R-correlator matches the singular point of the LO potential.
However, we emphasize that the singularity of the potential itself may not be problematic in the HAL QCD method; the analytical investigation on the HAL QCD method~\cite{Aoki:2021ahj} suggests that phase shifts can be correctly obtained  from potentials even with singularities  due to the zeros of the NBS wave functions. 
Nevertheless, we found that these singular potentials fail to provide correct  binding energies numerically,  indicating that it is difficult to obtain reliable results from such singular potentials. 
One of the possible solutions to overcome this difficulty is to find another potential which has no singularity but reproduces the same scattering amplitudes.

\subsection{Utilizing the mixed R-correlators in $8_{s(a)}$ channel}
\label{subsec:results_mixed}
As mentioned in the last paragraph of Sec.~\ref{sec:setups}, the $8_s$ and $8_a$ channels do not mix at the leading order in the chiral perturbation theory~\cite{Jido:2003cb}.
In addition, the meson-baryon systems in the two channels have the same interaction term at this order.
These two suggest that the low-energy spectra in both channels are the same, including the bound states. 
Considering this property, we assume that $8_s$ and $8_a$ are approximately degenerated at low energy. 
Then, R-correlators for the two channels are linear combinations of same NBS wave functions with different weights, which give different potentials but produce the same scattering amplitude. 
Moreover, the same property holds for the linear combination of the two R-correlators: $R^{\textrm{mix}}_{\alpha}(c;{\bf r},t) = R^{8_s}_{\alpha}({\bf r},t) - c R^{8_a}_{\alpha}({\bf r},t)$, where $R^{8_s}_{\alpha}({\bf r},t)$ and $R^{8_a}_{\alpha}({\bf r},t)$ are R-correlators in $8_s$ and $8_a$ representations, and $c$ is a constant. 
Therefore, setting $c$ such that $R^{\textrm{mix}}_{\alpha}(c;{\bf r},t)$ does not cross zero, we can obtain the potential without any singular behaviors, so that one can extract the binding energy in $8_{s(a)}$ channel more reliablly. 
In our study, we take $c=0.2$, $0.4$, $0.6$, and $0.8$, all of which are the values satisfying the above criterion.

\begin{figure}[t]
    \begin{center}
        \includegraphics[width=0.45\textwidth]{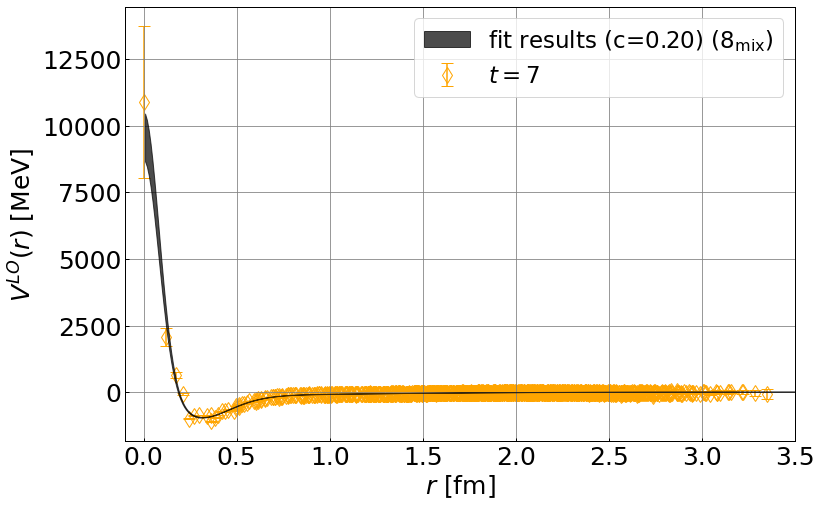}
        \includegraphics[width=0.45\textwidth]{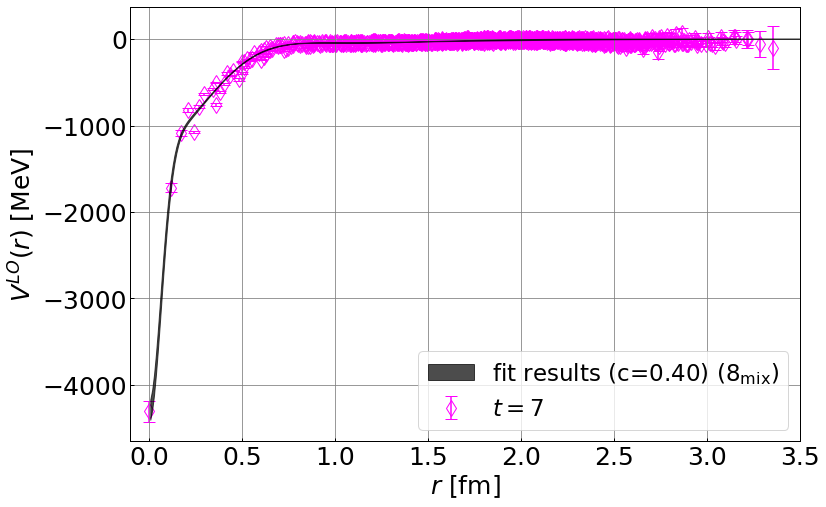}
        \includegraphics[width=0.45\textwidth]{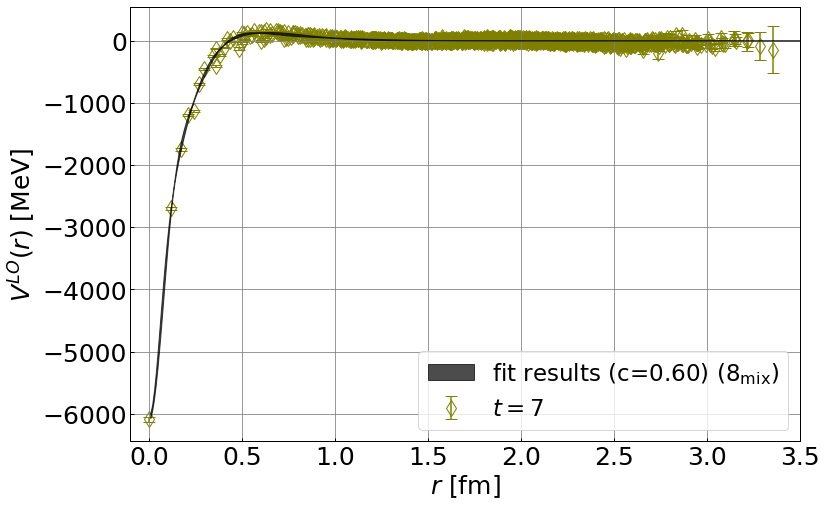}
        \includegraphics[width=0.45\textwidth]{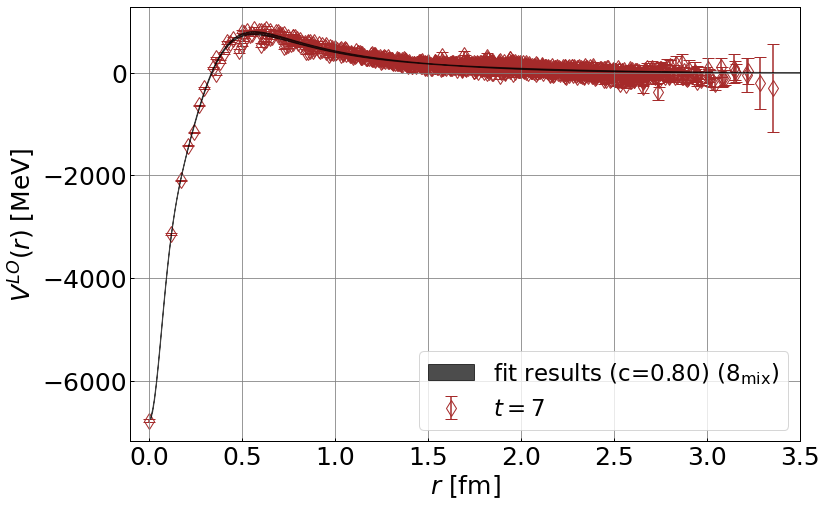}
	    \caption{LO potentials for $c=0.2$ (upper left), $0.4$ (upper right), $0.6$ (lower left), and $0.8$ (lower right) and their fit results (black bands). 
        }
	\label{fig:results_pot_mix}    
    \end{center}
\end{figure}

Fig.~\ref{fig:results_pot_mix} depicts the results of the LO potentials calculated from $R^{\textrm{mix}}_{\alpha}(c;{\bf r},t)$ in each $c$. 
All the potentials show strong attraction, which may creates bound states. 
On the other hand, they have different shapes: repulsive core with an attractive pocket for $c=0.2$, long-range attraction for $c=0.4$, attraction with a small bump at intermediate distances for $c=0.6$, and short-range attraction at short distances with a bigger bump at intermediate distances for $c=0.8$.

To extract the binding energy, we first fit the obtained potential by the Gaussian function, $V(r)=\sum^{N}_{i=1}a_{i}e^{-(r/b_{i})^2}$ with $N=5$ for $c=0.2$ and $N=4$ for others, and fit results are represented by black bands in Fig.~\ref{fig:results_pot_mix}.
We then solve the Schr\"{o}dinger equation using the Gaussian expansion method~\cite{Hiyama:2003cu}. 
Results of binding energies are listed in Table.~\ref{tab:BE_list}.
\begin{table}
    \centering     
    \begin{tabular}{c||c|c|c|c}
     $c$      & 
    $0.2$ &
    $0.4$ &
    $0.6$ &
    $0.8$ 
     \\ \hline
     $E_{\textrm{bind}}~\textrm{MeV}$&
    $179(4)$&
    $163(7)$&
    $132(13)$ &
    $99(15)$
    \end{tabular}        
    \caption{\label{tab:BE_list}
    The binding energy for each $c$ in $8_{s(a)}$ channel.
    }
\end{table}
We find that binding energies for different $c$ have similar values despite different shapes of potentials, which indicates that they generate almost identical dynamics
 at least in the low-energy regime.

Finally we take the binding energy for $c=0.4$ as the central value while systematic errors are estimated by values for other $c$.
We obtain the binding energy in the $8_{s(a)}$ channel  as $E^{8_{s(a)}}_{\textrm{bind}}=163(7)\smqty(+16 \\ -64)~\textrm{MeV}$, where the first and second parentheses correspond to the statistical and systematic errors, respectively.
The final result is consistent with the estimation from the two-point correlation function, $m^{+}_{8, B}+m_{M}-m^{-}_{8, B}=156(8)~\textrm{MeV}$, within errors, which indicates that approximation and assumption used in our analysis are reasonable.

\section{Summary}
\label{sec:summary}
We investigate the S-wave meson-baryon interaction potential in the time-dependent HAL QCD method for singlet and two octet channels, $8_{s}$ and $8_{a}$, in the flavor SU(3) limit, where the poles corresponding to $\Lambda(1405)$ have been predicted by the chiral unitary model. 
We use gauge configurations with the PS meson mass $m_M\approx 670~\textrm{MeV}$, in which each channel has at least one bound state. 
We compute the three-point functions with $\Lambda$-baryon source operators with zero momentum. 
To calculate the all-to-all propagators, we employ the conventional stochastic calculation combined with the covariantly approximated averaging. 
In this analysis, we perform the single-channel analysis for  singlet as well as $8_s$ and $8_a$ channels.
For all cases, the LO potential in each channel has a singular point due to the zero of the R-correlator, which makes it difficult to perform reliable analysis.  

Assuming that $8_s$ and $8_a$ are controlled by the same dynamics at low energy, we then combine  two octet R-correlators in order to avoid zeros.
The LO potential calculated from the mixed R-correlator has a strong attraction while its shapes depened on  coefficients.
The binding energy in the $8_{s(a)}$ channel is estimated as $E^{8_{s(a)}}_{\textrm{bind}}=163(7)\smqty(+16 \\ -64)~\textrm{MeV}$, consistent with that estimated from the two-point correlation function within errors. 
This indicates that our approximation and assumption are reasonable.

Nevertheless, it is difficult to draw a physical interpretation of the dynamics, since shapes of the obtained potentials depend on the parameter $c$, which is introduced 
to avoid zero in $R$-correlators. 
The appearance of zero in $R$-correlators for all channels may indicate strong non-local effects in interactions, which we ignore in this paper.  
Such non-locality, together with effects from the mixing and the breaking of the degeneracy for the bound states between $8_s$ and $8_a$ channels, should be examined in future.
Furthermore, we cannot apply the technique in Sec.~\ref{subsec:results_mixed} to the singlet channel since we have only one R-correlator.
Instead, a similar technique may be used in the singlet channel by employing the four-point correlation function with the meson-baryon source operator, which is left to our future studies.

\section{Acknowledgments}
We would like to thank the members of the HAL QCD Collaboration for fruitful discussions.
We also appreciate Profs. D.~ Jido and M.~Oka for their useful comments. 
We thank Prof. T.~Inoue and ILDG/JLDG~\cite{Amagasa:2015zwb} for providing us with gauge configurations used in this paper. 
We use the lattice QCD code of Bridge++~\cite{Ueda:2014rya, bridge++url} and its optimized version by Dr. I. Kanamori~\cite{Kanamori:2018hwh}.
Our research uses computational resources of Wisteria/BDEC-01 Odyssey (the University of Tokyo), provided by the Multidisciplinary Cooperative Research Program in the Center for Computational Sciences, University of Tsukuba.
K.~M. is supported in part by JST SPRING, Grant Number JPMJSP2110, by Grants-in-Aid for JSPS Fellows (Nos.\ JP22J14889, JP22KJ1870), and by JSPS KAKENHI Grant No.\ 22H04917.
This work is also supported in part by JPMXP1020230411.

\end{document}